\documentstyle[aps]{revtex}
\input{epsf}
\begin{document}
\flushbottom

\twocolumn

\def\thepage{\roman{page}}
\title{\vspace*{0.5in}
Ionization history of the Universe as a test for Super Heavy 
Dark Matter particles}

\author{A.G. Doroshkevich}
\address{Theoretical Astrophysics Center, Juliane Maries Vej
30, 2100 Copenhagen, {\O} Denmark;}
\author{P.D. Naselsky}
\address{Theoretical Astrophysics Center, Juliane Maries Vej
30, 2100 Copenhagen, {\O} Denmark;\\ Rostov State University, 
Zorge 5, 344090, Rostov-Don, Russia}
\maketitle
\newcommand{\bedm}{\begin{displaymath}}
\newcommand{\eedm}{\end{displaymath}}
\newcommand{\be}{\begin{equation}}
\newcommand{\ee}{\end{equation}} 
\flushbottom

\begin{abstract}
In this paper we  discuss the possible distortions of the ionization 
history of the universe caused by an injection of non-thermal energy 
due to decays of hypothetical Super Heavy Dark Matter (SHDM) particles. 
These particles are usually considered as a possible source of Ultra 
High Energy Cosmic Rays (UHECRs) in the framework of the Top-Down 
model. Estimates of fraction of energy of decays converted to the 
UV range show that, for suitable parameters of SHDM particles, 
the significant distortions of power spectra of CMB anisotropy 
appear. Comparison with the observed power spectrum allows to 
restrict some properties of the SHDM particles. These decays can 
also increase of about 5 -- 10 times the degree of ionization 
of hydrogen at redshifts $z\sim$ 10 -- 50 that essentially accelerates 
the formation of molecules $H_2$ and first stars during "dark ages". 

PACS number(s): 98.80.Cq, 95.35.+d, 97.60.Lf, 98.70.Vc.
\end{abstract}

\section {Introduction}

Two different approaches to investigation of the general physical 
properties of the cosmological expansion of the Universe are now 
in the fast progress. One of them is theoretical and experimental 
investigations of the Cosmic Microwave Background (CMB) anisotropy 
and polarization. Other is the investigation of possible manifestations 
of Super Heavy Dark Matter (SHDM) particles several kinds of which 
can be created at the period of inflation of the Universe.

The analysis of the CMB anisotropy and polarization is the "gold 
mine" for the determination of cosmological parameters such as 
fractions of baryons ($\Omega_b$), cold dark matter($\Omega_m$) 
and vaccua ($\Omega_\Lambda$), Hubble constant $h=H_0/100$ km/s/Mpc, 
power index $n$ of initial adiabatic (or isocurvature) perturbation, 
possible redshifts of reionization $z_{ion}$ and so on. Recent 
progress in this direction is based on unique information about 
the power spectrum of CMB anisotropy measured in the ground and 
balloon-borne experiments such as BOOMERANG \cite{1} and MAXIMA-1 
\cite{2}. It is generally believed that future high precision 
observations of the CMB anisotropy allow to test the most important 
predictions of the modern theories of inflation and will stimulate 
a further study of the very early Universe. 

On the other hand, recent observations of AGASA \cite{3}, Fly's Eye 
\cite{4} and Haverah Park \cite{5}, demonstrate the existence 
of the Ultra -- High Energy Cosmic Rays (UHECR) with energy above 
the Greisen-Zatsepin-Kuzmin (GZK) \cite{6,7} cutoff, $E_{GZK}\sim 
10^{20}$ eV, that is one of the most intriguing mysteries of the 
modern physics and astrophysics (see reviews \cite{8,9}). As was 
suggested by Berezinsky, Kachelrie$\ss$ \& Vilenkin \cite{10}, 
Kuzmin \& Rubakov \cite{11}, and Birkel \& Sarkar \cite{111}  
the formation of such UHECRs can be related to 
decays of the various kinds of SHDM X--particles with masses 
$M_X\geq 10^{12}$ GeV in the framework of the so-called Top-Down 
scenario of the UHECR creation. (Below we denote by $X$ all 
possible types of SHDM). 

As is commonly believed, decays of SHDM particles into the 
high energy protons, photons, electron-positron pairs and neutrinos 
occurs through the production of quark-antiquark pairs ($X\rightarrow
q,\overline{q}$), which rapidly hadronize, generate two jets and 
transform the energy into hadrons ($\omega_h\sim$5\%) and pions 
($1-\omega_h\sim$95\%) \cite{12}. It can be expected that later on 
the energy is transformed mainly to high energy photons and neutrinos. 
Other estimates of the photon, $\omega_{ph}$, and hadron, $\omega_h$, 
fractions produced by the decays ($\omega_{ph}\approx 2 - 3~\omega_h$), 
was obtained in \cite{111,112}. This means that, for both models of 
decays of SHDM particles with $10^{12}$GeV$<{M_X}<10^{19}$ GeV, the 
UHECR with energies $E>10^{20}$ eV are dominated by photons and 
neutrinos \cite{12,112}. 

However, recent observations Ave et al. [15]   
shown that above 
$10^{19}$eV less than 50\% of the primary cosmic rays can be photons. 
These observations demonstrate that probably only some part of the 
observed UHECR can be related to the decays of the SHDM particles, 
and more sensitive and refined methods must be used for further 
observational investigation of such X--particles. 

In this paper we discuss the possible distortions of the ionization 
history of the primeval plasma at redshifts $z\leq 10^3$ caused by 
the energy injection due to decays of the SHDM particles. For 
the model of unstable neutrinos similar analysis was performed in  
\cite{18}. Comparison of expected distortions with already available 
observational BOOMERANG and MAXIMA-1 data allows us to restrict more 
strongly the rate of decays of the possible SHDM particles and, at 
the same time, to refine evaluations of possible distortions of CMB 
anisotropy and polarization. We show also that, for reasonable 
parameters of the SHDM particles, their decays can increase of about 
5 -- 10 times the degree of ionization of hydrogen at redshifts 
$z\sim$ 10 -- 50 that essentially accelerates the formation of 
molecules $H_2$ and first stars during "dark ages". 

The paper is organized as  follows. In section II some information 
about of the SHDM particles is summarized. In section III the spectrum 
of UV radiation produced by the energy injection is found. In sections 
IV and V we discuss the delay of cosmological recombination and the 
hydrogen ionization history. In section VI we show that the expected 
distortions of the CMB spectrum (SZ--effect) are small. In section VII 
some restrictions of characteristics of the SHDM particles following 
from the available observations of the CMB anisotropy are discussed. 
Main results are summarized in section VIII.

\section{Expected flux of high energy photons}

The expected domination of products of decay of the X-particle by 
the high energy pions, neutrino and photons follows from quite general 
arguments. Probable energy losses of neutrinos are small \cite{8}, 
but at high redshifts the interaction of both high energy photons and 
hadrons with the CMB background leads to formation of electromagnetic 
cascades. At small redshifts the efficiency of this interaction 
decreases and the evolution of such photons depends upon unknown 
factors such as the extragalactic magnetic field and properties of 
radio background.  

Summarizing available information about the photon component of UHECRs 
Bhattacharjee and Sigl \cite{8} estimate the spectrum of injected 
photons for a decay of a single $X$-particle as follows:
\be
{dN_{inj}\over dE_{\gamma}}=\frac{0.6(2-\alpha)}{M_X}
\frac{f_{\pi}}{0.9}\left(\frac{2E_{\gamma}}
{M_{X}}\right)^{-\alpha},~ E_\gamma\leq {M_X\over 2}\,,
\label{eq1}
\ee
\bedm
E_{sum}=\int_0^{M_X/2} E_\gamma{dN_{inj}\over dE_{\gamma}}dE_{\gamma} 
= 0.15M_X(f_\pi/0.9)
\eedm
where $E_{sum}$ is the total energy of decay carried by photons, 
$f_{\pi}$ is the fraction of the total energy of the jet 
carried by pions (total pion fraction in terms of number of 
particles), $0<\alpha<2$ is the power index of the injected spectrum, 
$M_{X}$ is the mass of the SHDM particles. For the photons path length 
$l_{\gamma}\sim 1-10$Mpc at $E_{\gamma}\sim E_{GZK}$ in respect to the  
electron-positron par creation on the extragalactic radio background
(see {\cite{8,14}}) we get for the photon flux $j_{inj}(E_{\gamma})$ 
at the observed energy $E_{\gamma}$:
\be
j_{inj}(E_\gamma)\simeq\frac{1}{4\pi}l_{\gamma}(E_{\gamma})
\dot{n}_{X}{dN_{inj}\over dE_{\gamma}}\,,
\label{eq2}
\ee
where $\dot{n}_{X}$ is the decay rate of the $X$ - particles.
For the future calculation we will use the normalization of 
$j_{inj}(E_{\gamma})$ on the observed UHECR flux which corresponds 
to normalization of the decay rate $\dot{n}_{X}$ at present time, 
$t=t_u$, {\cite{8}}:
\be
\dot{n}_{X,0}\simeq 10^{-46}{\rm cm^{-3}s^{-1}}
M_{16}^{1-\alpha}\Theta_X\,,
\label{eq3}
\ee
\bedm
\Theta_X\approx\frac{10 {\rm Mpc}}{l_\gamma (E_{obs})}
\frac {E^2_{obs} j_{obs}(E_{obs})}
{\rm 1 eV cm^{-2} s^{-1}sr^{-1}}(2E_{16})^{\alpha-3/2}\frac{0.5}
{2-\alpha}\frac{0.9}{f_{\pi}}
\eedm
where $M_{16}=M_X/10^{16}$GeV, $E_{16}=E_{obs}/10^{16}$GeV, 
$E_{obs}$ and $j_{obs}(E_{obs})$ are the observed energy and flux
of the UHECR. Normalization (\ref{eq3}) does not depend on the nature 
of the $X$-particles. The precision achieved is about of order of 
magnitude. 

At $z\gg 1, t\ll t_u$ the decay rate depends on the physical nature 
of the X-- particles (see, for example, {\cite{8}}). To reconstruct 
the ionization history of the universe at redshifts $z\sim 10^3$ 
we will consider the simplest evolutionary model with:
\be
{dn_X(t)\over dt} + 3H(t)n_{X}(t)= -\frac{n_{X}(t)}{\tau_X}\,, 
\label{eq4}
\ee
\be
n_{X}(t)=n_{X,0}(1+z)^3\exp\left(\frac{-(t-t_u)}{\tau_X}\right)\,,
\label{eq5}
\ee
where $H(t)$ is the Hubble parameter, $z$ is a redshift, 
$t_u\sim H^{-1}(z=0)$ is the 
age of the universe, $n_X(t)$ and $\tau_X$ are the number density 
and life--time of the $X$-particles. 
In a general case, we can write 
\be
\dot{n}_{X}= {n_X(z)\over \tau_X} = \dot{n}_{X,0}(1+z)^3\Theta_\tau(z),
\quad \Theta_\tau(z)\geq 1\,,
\label{eq6}
\ee
For $\tau_X\geq t_u$, $\Theta_\tau\approx$1 and the decay rate, 
$\dot{n}_X$, varies only due to the general expansion. 

\section{Electromagnetic cascades at the period of 
the hydrogen recombination.}

To evaluate the distortions of the power spectra of CMB anisotropy 
and polarization we need firstly to consider the transformation of 
high energy injected particles to UV photons influenced directly the 
recombination process. 
The electromagnetic cascades are initiated by the ultra high energy 
jets and composed by photons, protons, electron- positrons and neutrino. 
At high redshifts, the cascades develops very rapidly via interaction 
with the CMB photons and pair creation ($\gamma_{UHECR}+\gamma_{CMB}
\rightarrow e^{+} + e^{-}$), proton-photon pair production ($p_{UHECR} 
+ \gamma_{CMB}\rightarrow p^{'} + \gamma{'}+e^{+} + e^{-}$), inverse 
Compton scattering ($e^{-}_{UHECR}+\gamma_{CMB}\rightarrow e^{'} + 
\gamma^{'}$), pair creation ($e^{-}_{UHECR}+\gamma_{CMB}\rightarrow 
e^{'}+ e^{-} + e^{+} + \gamma^{'}$), and, for neutrino, electron-
positron pair creation through the Z-resonance. As was shown by 
Berezinsky et al. \cite{15} and Protheroe et al. \cite{16}, these 
processes result in the universal normalized spectrum of the cascade 
with a primary energy $E_\gamma$:
\be 
N_\gamma(E)=\frac{E_\gamma E^{-2}}{2+\ln(E_c/E_a)}
\left\{
\begin{array}{cc}
\sqrt{E\over E_a}&E\leq E_a\cr
1&E_a\leq E\leq E_c\cr
0&E_c\leq E\cr
\end{array} 
\right.      
\label{eq1e}
\ee
\bedm
\int_0^{E_\gamma}EN_\gamma dE=E_\gamma\,,
\eedm
where $E_\gamma$ was introduced in (\ref{eq1}), $E_c\simeq 4.6\cdot 
10^4(1+z)^{-1}$GeV, $E_a=1.8\cdot 10^3(1+z)^{-1}$GeV. At the period 
of recombination $z\sim 10^3$ and for less redshifts both energies, 
$E_a$ and $E_c$, are larger then the limit of the electron-positron 
pair production $E_{e^{+},e^{-}}=2 m_e = 1$ MeV and the spectrum 
(\ref{eq1e}) describes both the energy distribution at $E\geq 
E_{e^+,e^-}$ and the injection of UV photons with $E\ll E_{e^+,e^-}$. 
However, the spectrum of these UV photons is distorted due to the 
interaction of photons with the hydrogen - helium plasma.  

In the range of less energy of photons, $E\leq 2m_e$, and at higher 
redshifts, $z\geq 10^3$, when equilibrium concentrations of $HI$, 
$HeI$ and $HeII$ are small and their influence is negligible, the 
evolution of the spectrum of ultraviolet photons, $N_{uv}(E,z)$, 
occurs due to the injection of new UV photons and their redshift 
and Compton scattering. It is described by the transport equation 
\cite{16}
\be
{\partial N_{uv}\over \partial z}-{3N_{uv}\over 1+z}+{\partial
\over \partial E}\left(N_{uv} {dE\over dz}\right) + {Q(E,z)\over 
(1+z)H} = 0,
\label{eqq1}
\ee
\bedm
{1+z\over E}{dE\over dz} = 1+ {c\sigma_Tn_e\over H(z)}
\left(\frac{E}{m_ec^2}\right)=1+\beta_\gamma(E,z)\,,
\eedm
\bedm
Q(E,t)= \dot{n}_X\int dE_\gamma N_\gamma(E,E_\gamma)
{dN_{inj}\over dE_\gamma}= 
\eedm
\bedm
0.15{f_\pi\over 0.9}\dot{n}_X N_\gamma(E,M_X)
\eedm
where the Hubble parameter is
\bedm
H(z)=H_0\sqrt{\Omega_m(1+z)^3+1-\Omega_m}\,,
\eedm
$\sigma_T$ is the Thomson  cross-section, $n_e\propto (1+z)^3$ 
is the number density of electrons and, so, $\beta_\gamma\propto 
(1+z)^{3/2}E$. Here $Q(E,t)$ is considered as an external source 
of UV radiation. 
 
The general solution of equation (\ref{eqq1}) is: 
\be
N_{uv}(z) =\int_z^{z_{mx}}{Q(x)\over H(x)}
{E^2(x)\over E^2(z)}\left({1+z\over 1+x}\right)^4{dx\over 1+x}, 
\label{eqw1}
\ee
\bedm
{E(x)\over E(z)}=\frac{1+x}{1+z}\left(1-{2\over 5}\beta_\gamma(E,z)
\left[\left({1+x\over 1+z}\right)^{5/2}-1\right]\right)^{-1},
\eedm
where the maximal redshift, $z_{mx}$, in (\ref{eqw1}) is defined 
by the condition $E(x)=2m_ec^2$.

As is seen from (\ref{eqw1}), the Compton scattering dominates for 
$E\gg 30$keV, when 
\be
\beta_\gamma(E,z) = 44{E\over m_ec^2}\sqrt{0.3\over\Omega_m}
{h\Omega_b\over 0.02}\left({1+z\over 10^3}\right)^{3/2}\gg 1,
\label{beta}
\ee
\be
N_{uv}(z)\propto \frac{\dot{n}_X(z)N_\gamma(E,M_X)}{H(z)
\beta_\gamma(E,z)}\propto {\sqrt{1+z}\Theta_\tau(z)\over E^{5/2}} ,
\label{eqw22}
\ee

For the most interesting energy range, $E\ll 30$keV,  
$\beta_\gamma(E,z)\ll 1$, we get again
\be
N_{uv}(E(z),z)\approx 0.1{f_\pi\over 0.9}\frac{\dot{n}_X(z)}{H(z)}
N_\gamma(E,M_X),
\label{eqw2}
\ee
that is $\sim 2/3$ of the photons produced by the spectrum of injection 
(\ref{eq1e}) in the same energy range. 

The energy density, $\Delta \epsilon$, produced by the decays at redshifts 
$z\geq 10^3$ near the energy of ionization of hydrogen and helium, $E
\simeq I_H$, is 
\be
\Delta\epsilon = \int_{I_H}^E EN_{uv}(E)dE\approx 
\kappa_H{\dot{n}_X M_X\over H(z)}\left(\sqrt{E\over I_H}-
1\right),
\label{eqeps1}
\ee
\bedm
\kappa_H\approx {0.21 f_\pi\over 2+\ln(E_c/E_a)}\sqrt{I_H\over E_a}
\approx 5f_\pi\cdot 10^{-6}\sqrt{1+z\over 10^3},
\eedm
\bedm
{\dot{n}_X M_X\over H(z)}\approx 25{eV\over cm^3}\left({z\over 10^3}
\right)^{3/2}\sqrt{0.15\over \Omega_mh^2}M_{16}^{2-\alpha}\Theta_\tau\Theta_X.
\eedm
where $I_H=13.54$ eV is the potential of ionization of hydrogen and $M_{16}$ 
is dimensionless mass of the SHDM particle introduced in (\ref{eq3}). 
For comparison, the energy density of the CMB radiation at 
$z=10^3,~~T_\gamma=2700(1+z)K$ and $E\geq I_H$ is 
\bedm
\Delta\epsilon_{bb}\approx 4\cdot 10^{-6}\left(\frac{1+z}{10^3}\right)
\exp\left[58.5\left(1-{10^3\over 1+z}\right)\right]{eV\over cm^3},
\eedm
that demonstrates the possible strong influence of decays for the 
recombination history. 

\section{Distortions of the CMB anisotropy and polarization.}

As is well known, at later stages of the standard model of recombination   
the rate of recombination depends upon the interaction of neutral 
hydrogen with numerous trapped Ly--$\alpha$ photons. This means that 
the external sources of both Ly--c and Ly--$\alpha$ photons with a 
suitable intensity delay the recombination process and shift the 
position of Doppler peaks. However, the injection of Ly-c photons 
after the recombination increases the ionization degree of hydrogen 
and leads to an additional suppression of the Doppler peaks due to 
the Thomson scattering of the CMB. For decays of neutrinos with 
$m_\nu\approx 27$eV this problem was considered in \cite{18} and a 
strong suppression of the CMB anisotropy as compared with the standard 
model was found. The same problem was discussed in \cite{17} for the 
action of arbitrary external sources of both Ly-c and Ly--$\alpha$ 
photons.  

The spectrum (\ref{eqw2}) gives a reasonable description of the cascade 
at higher redshifts, $z\geq 10^4$, when the concentrations of neutral 
hydrogen and helium are small. At redshifts $z\leq 10^3$ and for 
$E\simeq I_H$ the spectrum (\ref{eqw2}) is strongly distorted due 
to the reionization of hydrogen and helium, and the main part of 
energy (\ref{eqeps1}) is rapidly converted to the resonance lines, 
namely, $Ly-c=912{\AA}, ~\&~ 228{\AA}$ and $Ly-\alpha=1216{\AA}, 
~\&~ 304{\AA}$. However, as is seen from (\ref{eq1e}), (\ref{eqq1}) 
and (\ref{eqw2}), for the cascades generated by decays of the SHDM 
particles the expected numbers of Ly--c and Ly--$\alpha$ photons are 
comparable and, therefore, the action of Ly--c photons dominates. 

To estimate in the case the distortions of the CMB power spectrum, 
we can use the approach proposed in \cite{17} and write the rate of 
production of resonance and ionized photons, $\dot{n}_r$, as follows:
\be
 {dn_r\over dt} = {2\over 3}\int_{I_H}^E Q(E,z) dE = 
\varepsilon(z)\langle n_b(z)\rangle H(z) \,,
\label{eps}
\ee
\bedm
\varepsilon(z)\approx {0.13f_\pi\over 2+\ln(E_c/E_a)}{M_X\over 
\sqrt{I_HE_a}}{\dot{n}_X\over H(z)\langle n_b(z)\rangle}\,.
\eedm
The comparison (\ref{eqeps1}) and (\ref{eps}) shows that 
\bedm
\Delta\epsilon(4I_H,z)\simeq \varepsilon(z)I_H\langle n_b(z)\rangle\,.
\eedm
For a given $\varepsilon(z)$, the power spectra of CMB anisotropy, 
polarization and their cross-correlation can be found with the modified 
CMBFAST and RECFAST codes \cite{20}. These results confirm the dominant 
influence of directly injected Ly--c photons as compared with that of 
the Ly--$\alpha$ photons. 

For the mean number density of baryons  
\bedm
\langle n_b\rangle\approx 240{\Omega_bh^2\over 0.02}
\left({1+z\over 10^3}\right)^3 cm^{-3}\,,
\eedm
we have
\be
\varepsilon(z)\approx {1.3\cdot 10^{-5}f_\pi\over 1+z}
M_{16}^{2-\alpha}\Theta_{tot}\,,
\label{eps2}
\ee
\bedm
\Theta_{tot}=\sqrt{0.15\over\Omega_mh^2}\left({0.02\over \Omega_bh^2}
\right)\Theta_X\Theta_\tau(z)\,.
\eedm

Relations (\ref{eps}, \ref{eps2}) link the rate of injection of UV 
radiation and distortions of the power spectra of CMB anisotropy 
with the mass and life--time of X-particles, $M_X ~\&~\tau_X$ or 
$\Theta_\tau$, and the spectral index, $\alpha$, that allows to 
restrict latest ones using the available observations of CMB 
anisotropy. These restrictions will be considered in Sec. VII. 

For the models introduced in Sec. II, it can be expected that 
the function $\Theta_\tau(z)\sim const.\geq$1 and $\varepsilon(z)\propto 
(1+z)^{-1}$ at least at $z\geq$10--50 instead of the $\varepsilon=const.$ 
considered in \cite{17}. Models with $\Theta_\tau\propto (1+z)^\nu$, 
$\nu\geq -1$, and, in particular, with $\varepsilon(z)=const.$ can be 
considered in context of other kinds of the SHDM particles \cite{8}.

\begin{figure}
\centering
\epsfxsize=8 cm
\epsfbox{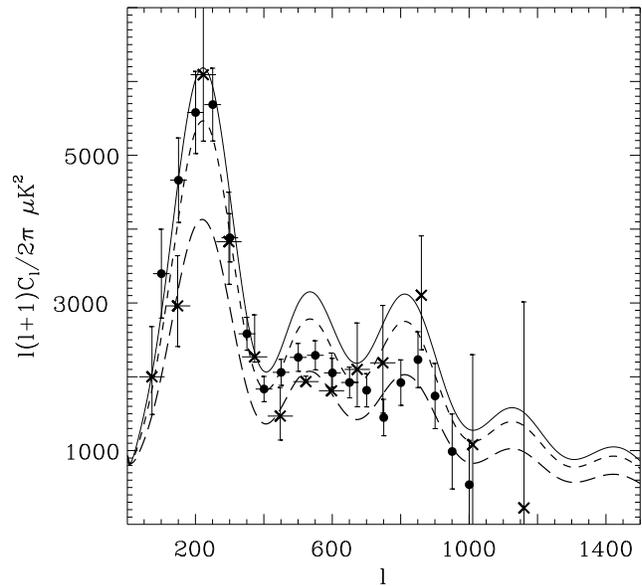}
\vspace{0.5cm}
\caption{
The power spectra of CMB anisotropy, $l(l+1)C_l/4\pi$, vs. 
$l$ are plotted for the standard model, ($\varepsilon= 0$, solid 
line), and for models with $\varepsilon= 2/(1+z)$, (dashed line),
and $\varepsilon= 3/(1+z)$, (long dashed line). Observational data 
are plotted by points (BOOMERANG) and crosses (MAXIMA-1). 
}
\end{figure}

Here we consider the cosmological 
model with $\Omega_b h^2=0.02$, $\Omega_m=0.3$, $\Omega_{\lambda}=0.653$, 
$h=0.65$ and the Harrison-Zel'dovich primordial power spectrum of initial 
adiabatic perturbations ($n=1$) without a contribution of gravitation 
waves and a later reionization caused by the galaxy formation.
For the standard model with $\varepsilon=0$ and for two models with 
$\varepsilon(z)= 2/(1+z)~\&~ 3/(1+z)$, the power spectra of anisotropy 
and polarization, $l(l+1)C(l)/2\pi$ and $l(l+1)C_{pl}(l)/2\pi$, are 
plotted in Figs. 1, 2. In Fig. 3 the cross correlation of anisotropy 
and polarization is also presented. 

As is seen from Fig. 1, the power spectrum of anisotropy is very 
sensitive to the influence of additional UV background that in turn 
restricts the intensity of UV radiation and characteristics of 
X-particles. Thus, this influence becomes negligible for $\varepsilon
\leq (1+z)^{-1}$ while for $\varepsilon\geq 3/(1+z)$ the CMB scattering 
at redshifts $z\leq 10^3$ results in an essential suppression of all 
Doppler peaks. For $\varepsilon\sim 2/(1+z)$, the expected power 
spectrum is well consistent with available observational data \cite{1,2}. 

\begin{figure}
\centering
\epsfxsize=8 cm
\epsfbox{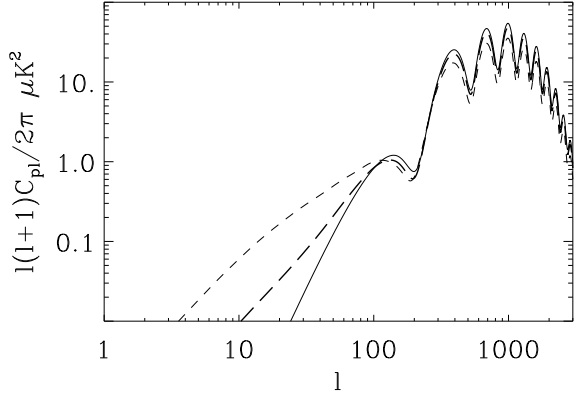}
\vspace{0.5cm}
\caption{The CMB polarization vs. $l$ for $\varepsilon=0$ (solid line), 
$\varepsilon= 2/(1+z)$ (dashed line) and $\varepsilon= 
3/(1+z)$ (long dashed line). 
} 
\end{figure}

The main suppression of the CMB anisotropy occurs due to the partial 
reionization of the hydrogen at redshifts $z\sim$ 600 -- 200. This 
effect depends upon the redshift variations of the rate of decay of 
the SHDM particles at the same redshifts. The shift of the Doppler 
peaks caused by the delay of the hydrogen recombination at $z\sim 
10^3$ is small ($\leq$ 1\%) for all $\varepsilon$ under consideration. 
This shift is not seen in the available observations but perhaps can 
be revealed with more detailed measurements of the MAP and PLANCK 
space missions.   

\section{The hydrogen ionization history}

The main distortions of the power spectrum 
of anisotropy are the suppressions of the Doppler peaks due to reionization 
of the Universe at redshifts $z<~ 10^3$. At these redshifts the degree of 
ionization of hydrogen is small, all Ly--c photons are rapidly absorbed, 
and the fraction of ionized hydrogen atoms can be roughly estimated from 
the equilibrium equation which describes the conservation of number of 
electrons and Ly--c photons together, 
\be
{d x_H\over dt}=\alpha_{rec}^*\langle n_b\rangle x_e^2 - 
x_H\varepsilon H(z)=0\,,
\label{xe}
\ee
\bedm
\alpha_{rec}^{*}\simeq 4\cdot 10^{-13}\left({T\over 10^4 K}
\right)^{-0.6}{cm^3\over s},~ T\approx 300K\left({1+z\over 100}\right)^2.
\eedm
Here $x_H$ and $x_e=1-x_H$ are the fractions of neutral hydrogen and 
electrons, respectively, $\alpha_{rec}^{*}$ is the recombination 
coefficient for states with the principle quantum number $n\geq$2, 
$T$ is the temperature of hydrogen under the condition of small 
ionization at $(1+z)\leq 100$ and $\langle n_b\rangle$ is the mean 
number density of baryons (\ref{eps}). For simplicity, we neglected 
here the contribution of helium. Numerically, we have from (\ref{xe})
\be
x_e^2\simeq\frac{\varepsilon H(z)}{\alpha_{rec}^*\langle n_b
\rangle}\sim 10^{-4}\varepsilon(0)\left({100\over 1+z}\right)^{5/4}
\Theta_{tot},
\label{xe1}
\ee
that essentially exceeds the standard estimates of 'frozen' ionization 
degree $x_e\sim 10^{-3}$ \cite{17}. 

\begin{figure}
\centering
\epsfxsize=8 cm
\epsfbox{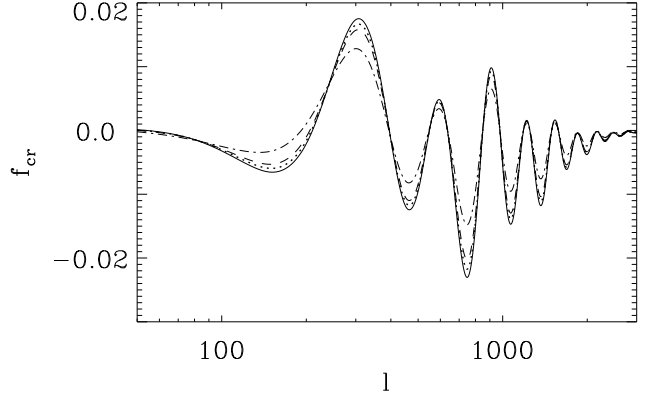}
\vspace{0.5cm}
\caption{The cross-correlation function of anisotropy and polarization 
for the same three values of $\varepsilon$.} 
\end{figure}

The optical depth achieved due to 
the reionization can be estimated as follows:
\be
\tau_D=c\sigma_T\int_0^z {\langle n_b(z)\rangle x_e(z)\over 
(1+z)H(z)}dz\,,
\label{tau_T}
\ee
\bedm
\approx 2.6\int_0^{z/100} \sqrt{y}x_e(y)dy\sqrt{0.3\over\Omega_m}
{\Omega_bh\over 0.02}\propto\sqrt{\varepsilon(0)}\,,
\eedm
where $y=(1+z)/100$ and again $c$ and $\sigma_T$ are the speed of 
light and the Thomson cross--section. As is seen from (\ref{xe1}, 
\ref{tau_T}), the optical depth is sensitive to the redshift 
$z\sim$600 -- 200 where the injected energy increases the ionization 
degree as compared with the standard model with $\epsilon=0$. 

For a given $\varepsilon(z)$, the ionization history can be restored 
more accurately with the modified RECFAST code \cite{20}. 
In Fig 4. the fraction of ionized hydrogen, $1-x_H$, is plotted 
versus redshift for the standard model and three values of 
$\varepsilon$. 

The fraction of ionized hydrogen, $1-x_H$, drops up to $\sim 10^{-3}$ 
at redshifts $z\sim 600$ and progressively increases at less $z$ up 
to $1-x_H\rightarrow 0.01 - 0.1$ at $z\leq 10$. Even for $\varepsilon=
0.3/(1+z)$ when the distortion of recombination is negligible, the 
ionization degree of hydrogen at $z\leq 50$ exceeds the standard one 
of about 5 -- 10 times that essentially accelerates the formation 
of molecules $H_2$ and first stars. At the same time, these results 
indicate that the UV flux generated by decays of X--particles is 
small as compared with the actually observed at $E=I_H$ and $z\sim$ 
3 flux \cite{19} 
\be
J\approx (1\pm 0.5)\cdot10^{-21}erg~cm^{-2}s^{-1}st^{-1}Hz^{-1}\,.
\label{eqjobs}
\ee 
This flux is mainly produced by an activity of quasars and galaxies. 

\begin{figure}
\centering
\epsfxsize=8 cm
\epsfbox{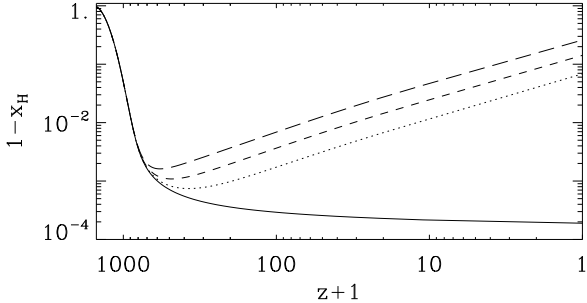}
\vspace{0.5cm}
\caption{The fractions of ionized hydrogen, $1-x_H$, vs. $1+z$ are 
plotted for $\varepsilon=0$ (solid line), $\varepsilon=
0.3/(1+z)$ (dot line), $\varepsilon=1/(1+z)$ (dashed line),
and $\varepsilon=3/(1+z)$ (long dashed line).
}
\end{figure}

\section{Estimates of the SZ effect}

Restrictions of the energy injection obtained above allow to 
estimate also the distortions of the CMB spectrum  -- the SZ 
effect. These distortions are generated due to the reheating 
of the hydrogen and a subsequent fast transmission of the thermalized 
energy to the CMB due to the inverse Compton cooling of hydrogen 
plasma. This cooling is very effective at $z\geq$ 15 and becomes 
negligible at less redshifts.

As was found above for the model under consideration, the rate of 
energy injection is restricted by the condition $\varepsilon\leq 
3/(1+z)$. Using the general theory of the SZ--effect and assuming 
that the energy injected in range $I_H\leq E\leq E_{mx}(z)$ is 
thermalized we get for the rate of energy injection per one baryon
\bedm
{d{\cal E}\over dt}\approx \int_{I_H}^{E_{mx}}{EQ(E,z)\over\langle 
n_b\rangle} dE = H(z)\varepsilon(z)\sqrt{E_{mx} I_H} 
\,,
\eedm
and for the difference of the electron and CMB temperature  
\bedm
k(T_e-T_r)={2\over 3}{\tau_c\over x_e}\dot{\cal E},\quad \tau_c = 
{3m_ec\over 8\sigma_T\epsilon_{rad}}\,,
\eedm
where $\tau_c$ is the characteristic time of the inverse compton 
cooling and $\epsilon_{rad}(z)\approx 0.25(1+z)^4$eV/cm$^3$ is the 
energy density of relic radiation. For the $y$--parameter we have
\be
y\approx {1\over 4}\int_{15}^\infty {\varepsilon(z)dz\over 1+z} 
{\sqrt{E_{mx}I_H}\langle n_b\rangle\over \epsilon_{rad}(z)}\approx 
\label{y1}
\ee
\bedm
0.7\cdot 10^{-5}\varepsilon(0){\Omega_bh^2\over 0.02}
\int_{15}^\infty {dz\over(1+z)^3}\sqrt{E_{mx}(z)\over I_H}\,,
\eedm
\bedm
y \leq 1.5\cdot 10^{-8}\varepsilon(0)\sqrt{E_{mx}\over I_H}
\left({15\over 1+z}\right)^2{\Omega_bh^2\over 0.02}\,.
\eedm

This result demonstrates that for any reasonable $E_{mx}(z)\leq 
m_ec^2$ the $y$--parameter is negligible as compared with the 
observed upper limit $y\leq 1.5\cdot 10^{-5}$ \cite{20}. 

\section{Restrictions of characteristics of the SHDM particles}

As was noted above, restrictions of the rate of injection of UV 
radiation due to distortions of the power spectra of CMB anisotropy 
$\varepsilon(z)\leq 3/(1+z)$ obtained in Sec. IV allow to restrict 
using (\ref{eps2}) the main characteristics of the SHDM particles, 
namely, the mass and life--time of X-particles, $M_X ~\&~\tau_X$ 
or $\Theta_\tau$, and the spectral index, $\alpha$. 

Here we do not specify the kind of SHDM particles and their 
properties as for the small life -- time the concentration of such 
particles at $z=0$ can be small and they cannot be detected as UHECRs. 
Non the less, if their concentration at $z=0$ is still significant 
then such particles can be linked to some fraction of observed UHECRs 
and their properties can be specified using the observational 
information about the UHECRs (see detailed discussion in \cite{8}). 
  
\begin{figure}
\centering
\vspace{0.2cm}
\epsfxsize=8 cm
\epsfbox{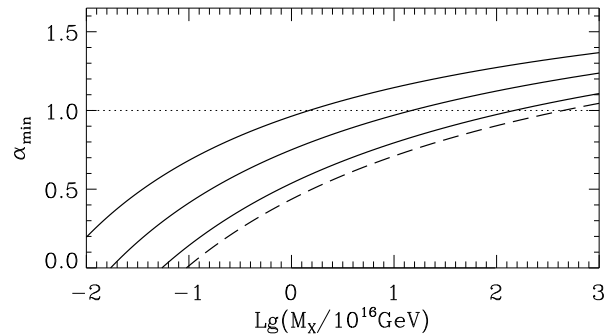}
\vspace{-0.4cm}
\caption{The spectral index, $\alpha_{min}$, vs. the mass of decaying 
X-particle, $M_X$, plotted for $\Theta_{tot}^*=1$ and $\varepsilon=
1, ~0.1, ~\&~ 0.01$ (solid lines), and $\varepsilon=3$ (long dashed 
line).
}
\end{figure}

In particular, the upper limit of $\varepsilon(z)\leq 3/(1+z)$ allows 
to link the mass of the X-particle with their life -- time and the 
power index of the spectrum of decay.  For $E_{obs}=10^{11}$GeV 
and $\lg\Theta_X\approx 4.7(1.5-\alpha)$ \cite{8}, the expression 
(\ref{eps2}) can be rewritten as follows:
\be
\alpha\geq\alpha_{min}\approx 2-{7.35+\lg(\varepsilon(z))-
\lg\Theta_{tot}^*\over 4.7+\lg(M_{16})},
\label{alpha-m}
\ee
\bedm
\Theta_{tot}^*=\Theta_\tau\sqrt{0.15\over\Omega_mh^2}{0.02\over
\Omega_bh^2}{\Theta_X\over 10^{4.7(\alpha-1.5)}}\,.
\eedm
Factors $\Theta_X$ and $\Theta_\tau$ were introduced in (\ref{eq3}) 
and (\ref{eq6}).

Restrictions obtained above relate mainly to the energy injection 
at redshifts $z\sim$100 -- 600 when $\varepsilon\sim (0.5 - 3)
\cdot 10^{-2}$. 
For long lived particles with $\Theta_\tau\approx 1$ and for four 
values $\varepsilon$ and $\Theta_{tot}^*=1$, the function 
$\alpha_{min}(M_X)$ is plotted in Fig. 5. This Fig. demonstrates 
that values $\alpha\leq$1 is consistent with observational restrictions 
only for $M_X\leq 10^{17}$GeV and/or $\Theta_{tot}^*\leq 1$. On the 
other hand, for the most popular value $\alpha\sim$1.5 the measurable 
distortions of power spectrum of CMB anisotropy appear for 
\be
\Theta_{tot}^*\leq \sqrt{10^6\over M_{16}}{300\over 1+z}
\label{alpha-1.5}
\ee
and for $z=300$ and $\Theta_{tot}^*=\Theta_\tau\approx\exp(t_u/
\tau_X)$ we get for the threshold life--time, $\tau_X$,
\be
t_u/\tau_X\leq 0.5\ln(10^6/M_{16})\,, 
\label{tau-1.5}
\ee
that restrict the life--time of the SHDM particles. 

For other kinds of SHDM particles with $\Theta_\tau=(t_u/t)^p
\propto (1+z)^{1.5p}$, $\varepsilon\propto (1+z)^{1.5p-1}$ results  
crucially depend upon the exponent $p$ and restrictions of both the 
energy injection and characteristics of the SHDM discussed above 
must be corrected. In particular, for $p=2/3$, the model discussed 
in \cite{17} with the injection of Ly--c photons and $\varepsilon
\sim (1 - 3)\cdot 10^{-3}$ can be used. 

These estimates demonstrate that the observations of power spectrum 
of CMB anisotropy can be used to restrict the possible properties of 
the SHDM particles.

\section{Summary and discussion}

 In the paper we consider some observational consequences of the 
energy injection due to possible decays of SHDM particles. As was 
noted in Introduction, many kinds of such particles are now 
discussed, in particular, in context of production of UHECRs 
(see, e.g., \cite{8,9,10,11,111}). 

We show that the energy of decay is transformed to UV range with 
a reasonable efficiency $\kappa\approx 10^{-5}$ (\ref{eqeps1}) that, 
for suitable mass and rate of decay of SHDM particles, can delays 
the hydrogen recombination, increase the ionization degree at less 
redshifts and provides the observed distortions of the CMB anisotropy. 
The possible action of these factors must be taken into account in 
interpretation of measured anisotropy together with usually considered 
main cosmological parameters. 

Let us note, that for all Top--Down models of the UHECRs the number 
of injected Ly--$\alpha$ and Ly--c photons are comparable, the delay 
of recombination is small and the main effects are the partial 
ionization of hydrogen and damping of the CMB anisotropy due to 
the Compton scattering of CMB at redshifts $z\leq$ 600. The action 
of the same decays can increase of about 5 -- 10 times the degree 
of hydrogen ionization at redshifts $z\leq$ 50 and accelerate the 
formation of $H_2$ molecules and first stars during "dark ages". 

We show that already available observations of the CMB anisotropy 
\cite{1,2} restrict the rate of injection of UV radiation and, so, 
the characteristics of discussed SHDM particles. For simple models 
these restrictions are presented in Sec. VI . These estimates can be 
also repeated for more exotic and refined models of the SHDM particle.

\section*{Acknowledgment}
Authors are grateful to V. Berezinsky, P. Blasi, P. Chardonet, 
M. Demianski, I. Novikov and I. Tkachev for discussions and help 
during the preparation of the paper. We are grateful to S.Sarkar 
and anonymous referee for the useful comments. 
This paper was supported in part by Danmarks Grundforskningsfond 
through its support for the establishment of the Theoretical 
Astrophysics Center, by grants RFBR 17625 and INTAS 97-1192.

\end{document}